# Comparative analysis of ADTCP and M-ADTCP: Congestion Control Techniques for improving TCP performance over Ad-hoc Networks.

Sreenivasa B.C [1], G.C. Bhanu Prakash [2], K.V. Ramakrishnan [3]

[1]Sr. Lecturer Department of Computer Science and Engineering
Research Scholar, Anna University Coimbatore.
[2]Professor Department of Computer Science and Engineering,
Sir M Visvesvaraya Institute of Technology- Bangalore, INDIA
[3] Visiting professor Department of Computer Science and Engineering,
CMRIT- Bangalore, INDIA
Email: srinivasbc@rediffmail.com

***ABSTRACT***

*Identifying the occurrence of congestion in a Mobile Ad-hoc Network (MANET) is a major task. The inbuilt congestion control techniques of existing Transmission Control Protocol (TCP) designed for wired networks do not handle the unique properties of shared wireless multi-hop link. There are several approaches proposed for detecting and overcoming the congestion in the mobile ad-hoc network. In this paper we present a Modified AD-hoc Transmission Control Protocol (M-ADTCP) method where the receiver detects the probable current network status and transmits this information to the sender as feedback. The sender behavior is altered appropriately. The proposed technique is also compatible with standard TCP.*

## *1. Introduction*

Mobile Ad-hoc Networks (MANET) do not have a fixed infrastructure. MANETs uses standard IEEE 802.11 MAC. In ad-hoc network each node (Mobile device) acts as a router, which helps in forwarding packets from a source to destination. MANETs are suitable in situations where fixed infrastructure is unavailable such as Military war fields, disaster relief, sensor networks, Wireless mesh network etc.

TCP congestion control is very much suitable for Internet, whereas for MANETs the same TCP is not suitable due to some of the specific properties like node mobility and shared wireless multi-hop channel. A slow delivery and packet loss occurs due to node mobility and unreliable shared medium. The delay in the packet delivery or packet losses is due to route change should not be misread as congestion.

In Internet when congestion occurs it is normally concentrated on a single router, whereas, due to the shared medium of the MANET congestion will not overload the mobile nodes but has an effect on the entire coverage area. The changes in the routing of the packet might lead to packet losses which is not caused due to congestion in the network should not be erroneously misinterpreted as TCP congestion. This can lead to wrong reactions of TCP congestion control. Further-more, monitoring packet losses is much harder, because of their varying transmission time and round trip time.

Many devices in ad-hoc network, sharing a common resource (i.e., media) compete for link bandwidth, which leads to network overload. When more data packet arrives at the router, the un-serviced packet gets dropped. These dropped packets would have consumed most of the network resources. The lost packets have to be retransmitted, which in turn leads to pumping of more packets into the network, resulting in degradation of network throughput and leading to congestion. To avoid congestion and network overload each sender has to adjust its data sending rate and window size.

A lot of research is being carried out in the area of congestion control, routing of packets, modification of standard TCP protocol, designing of new routing protocol, etc. in MANET,.

In OSI reference model, congestion control is the responsibility of the transport layer. The combination of congestion control and reliability features in TCP, allows congestion control management without the information about congestion status of the network. A proper mechanism is to be adopted to avoid congestion collapse of the MANET, which lead to the modification of TCP congestion mechanism [1]. The modified TCP should provide error and flow control. Flow control guarantees that the sender does not flood out the receiver by sending data at a rate faster than the receiver can process. It should also provide reliable end-to-end transmission of data over MANETs. The modified TCP should be capable of providing full-duplex, reliable and byte-stream services to the application programs.

The rest of this paper is structured as follows: Section 2 reviews related work, Section 3 discusses about modification made to ADTCP technique. Performance analysis of M-ADTCP and Comparison with ADTCP is given in Section 4, The Conclusion is given in Section 5.

## 2. Related work

A suitable congestion control technique for MANET is considered as an important issue. Some of the congestion related issues like throughput degradation and flow fairness are initiated from Media Access Control (MAC), routing and transport layer as discussed in [2][3][4][5]. Several papers have addressed and provided suitable solutions to overcome these problems.

A wireless link is prone to random packet losses unlike wired network. These losses affect the transport protocols performance, if they are wrongly interpreted as congestion induced by dropped packets. The link layer provides single hop reliability in 802.11 MAC protocol. The packets are dropped by link layer, only after maximum transmission attempts. This occurs when either a link is lost or due to packet collision. This section mainly deals with different approaches for congestion control in wireless ad-hoc network.

### Cross-layer congestion control ($C^3$TCP)

In this mechanism two network metrics, bandwidth and delay are measured between source and destination by cumulating intermediate hop measurements. This scheme is proposed by Kliazovich et al [6] and is similar to Rate-Based Congestion Control (RBCC) proposed by Zhai et al. [7]. In this technique a feedback field where the collected information at each intermediate node is stored and added to the link layer header. When ACK is generated at destination node, the feedback information of the data packet is transmitted to the sender. This information is used to modify receiver window field in ACK. It is also used to modify the windows size of the sender, which is located above TCP stack as an additional module. All C3TCP logic is part of additional protocol module without disturbing original TCP.

### TCP with Adaptive Pacing (TCP-AP)

ElRakabawy et al.[1] proposed a technique TCP-AP. This technique adopts an end-to-end approach for congestion control unlike C3TCP and RBCC. TCP-AP is a combination of both window and rate based approach. TCP is added with rate based mechanism to avoid large burst of packets.

In this technique the author proposes 4 hops propagation delay as a metric, measured using RTT of the packets. This is assumed as any interference if it happens within 4 hops. This propagation delay is the time elapsed between the transmissions of packet by source node to the receiving node 4 hops downstream. In order to estimate minimum time elapsed between successive packets, an additional metric ie., the coefficient of variation of RTT samples is used.

### TCP with Restricted Congestion Window Enlargement (TCP/RCWE).

Gunes¸ and Vlahovic [8] proposed a technique based on Explicit Link Failure Notification (ELFN) mechanism. In this technique the value of Retransmission Time Out (RTO) is observed randomly. The congestion window size is increased if the RTO value remains constant or decreases. If the RTO value increases the congestion window size is unaltered. The author has conducted NS-2 simulation using RCWE and reported lower packet losses and higher throughput due to smaller congestion window. The actual performance improvement due to ELFN is could not, as simulations are based on standard TCP without ELFN.

### Ad-hoc TCP (ADTCP)

ADTCP proposed by Fu et al.,[9] uses two metrics, *inter-packet delay difference* and *short-term throughput* to detect network congestion. The time elapsed between two successive packets and the throughputs in certain time interval in the immediate past are defined as inter-packet delay difference and short-term throughput respectively. When congestion occurs, *inter-packet delay difference increases, short-term throughput* decreases. To detect the channel error and route change, this technique uses *out of order packet arrival* and *packet loss ratio*. In ADTCP the accrued information at the receiver is sent as a feedback to the sender.

### Edge-based approach

In this technique, Measured RTT is used by De Oliveira et al.[10][11] to differentiate between medium losses and congestion. In this approach TCP congestion control reaction is avoided, when medium loss is detected. If packet is not received for a longer duration of time, it is identified as a route failure. To establish a new route the packets are transmitted at regular intervals. A fuzzy logic based approach is used to distinguish congestion from medium related losses
.

A single node can cause collapse of the entire network accidentally because of limited bandwidth in MANET. Thus the effect of single traffic flow may cause major unfairness in flow control. Multi-hop wireless network is more prone to overload related problems as compared to wired networks. Hence a suitable congestion control for a MANET is necessary for a stable and satisfactory network performance.

# 3. M-ADTCP

TCP has been predominantly used as transport protocol in the wired Internet to deliver data; consequently, numerous Internet applications have been developed to run over TCP. However, as explained earlier, TCP do not work satisfactorily in ad-hoc networks.

## 3.1 Concept

TCP in an ad-hoc network should be capable of handling disconnection and reconnection, packet out of order delivery in case of route change and errors due to node mobility in addition to congestion control.

In our technique we have adapted end-to-end measurement without considering explicit network notification mechanism. The measurements carried out at the receiver for every time interval α, are used to compute the status of the network to identify congestion related parameters. These parameters are carefully observed to initiate appropriate congestion control action for next iteration.

In MANET, the false congestion detections and notifications occur due to noise associated with measurements made at end hosts. Round-Trip Time (RTT) or packet inter-arrival time is not the ideal metric for detection of congestion, as the measured data is noisy [14]. The probability of false congestion detection is more in uncongested MANET, when only a single metric measurement is used. This leads to low TCP throughput.

In this paper we have proposed 4 metrics for identifying congestion. These metrics enable us to reduce noise in the measured data, thereby reducing probability of false congestion identification.

In M-ADTCP, the following metrics are devised to detect congestion,
IDD (Inter Delay Difference),
STT (Short Term Throughput),
POR (Packet Out of delivery Rate)

In congested state all these four metrics exhibit unique characteristics. The measurements made during the uncongested state mainly depend on prevailing network conditions and independent of noise measurement. Use of all these metrics reduces the false detection of congestion in the network.

## 3.2 Computation of End-to-End metrics

M-ADTCP sender uses the Round-Trip Time (RTT) to calculate the retransmission timeout. In ad-hoc networks packet delay is not only due to queue length, but also depends on other factors like random packet loss, changes in the route, MAC layer contention, etc. The process of computation of each metric in detail follows.

**Inter-packet Delay Difference (IDD):**

IDD indicates the congestion level along the path for each time interval. The receiver computes delay using Eq.1 for each packet received. The average IDD is computed for every time interval α (≈ 0.9s) to ascertain status of network.

$$IDD_i = (A_{i+1} - A_i) - (S_{i+1} - S_i) \quad (1)$$

$$IDD_{[T,T+\alpha]} = avg(IDD_{(i)}) \quad (2)$$

the time interval T to T+α

Where,

IDD: Inter Packet Delay Difference
$A_{i+1}$ : Arrival time of packet i+1
$A_i$ : Arrival time of packet i
$S_{i+1}$ : Sending time of packet i+1
$S_i$ : Sending time of packet i

**Algorithm for Calculation of IDD**

```
ComputeIDD(st, ed)//start and end packet number
repeat  // i is the packet number
        if(snd[i] and rcvd[i] and rcvd[i+1] and snd[i+1])
           idd+= (rcvd[i+1]-rcvd[i])-(snd[i+1]-snd[i])
 until (i<ed)
idd=idd/(ed-st+1)
```

In the algorithm, the function "ComputeIDD" calculates IDD for each interval. The arguments to this function are the start and end packets for a particular interval. The data structures rcvd[] and snd[] contain the times at which each packet is received and sent respectively. These arrays are indexed by the sequence number of the packets.

**Short-Term Throughput (STT)**:

The STT computation is independent of out-of-order packet delivery. The frequent changes in path selection do not influence STT calculation. The equation for computation of short-term throughput is as follows:

$$STT_{(i)}= N_p(T_i)/T_i \quad (3)$$

$$STT_{[T,T+\alpha]} =avg(STT_{(i)}) \quad \text{for each time interval T to T+}\alpha \quad (4)$$

$N_p(T_i)$: Total number of M-ADTCP packets received in the time interval $T_i$.
$STT_{[T,T+\alpha]}$ is the average Short-Term Throughput in the time interval [T,T+α] where α is 0.9 sec.

**Packet Out-of-order delivery Ratio (POR):**

If the difference between sequence-numbers of a packet received and that of previous packet is > 1 then current packet is counted as out-of-order in a single hop wireless network. In case of route change in multi-hop wireless network a packet may take a different path leading to out-of-order delivery. This case is not considered for POR computation. The equation for computation of POR is as follows:

$$POR_{(i)} = N_{po}(T_i)/ N_p(T_i) \quad (5)$$

$$POR_{[T,T+\alpha]}=avg(POR_{(i)}) \quad \text{for each time interval T to T+}\alpha \quad (6)$$

Where ,

$N_{po}(T_i)$ Total number of out-of-order packets during time interval $T_i$,
$N_p(T_i)$ Total number of packets received in the time interval $T_i$.

$POR_{[T,T+\alpha]}$ is the average Packet Out-of-order delivery Ratio in the time interval [T,T+α] where α is 0.9 sec.

# 4. Performance Evaluation and Results Analysis

We have implemented M-ADTCP and ADTCP technique using Network Simulator NS-2 Version 2.33.

## A. Simulation Parameters

The network consists of 5 nodes in a 670m x 670m square field. The MAC layer is configured to IEEE 802.11. Interface queue at MAC layer is set to default number of packets. The nominal bit rate is 2 Mbps and transmission range is 250 m. The Two Ray Ground model is used with maximum node speed of 4m/s. DSR is used as a routing protocol. The simulation time is 150 seconds. Constant Bit Rate (CBR) traffic is introduced at a rate of 1Mbps between node(0) and node(3) and at a rate of 0.75Mbps between node(3) and node(4) with packet size of 1500bytes. FTP traffic is introduced between node (1) and node (2) with default packet size and M-ADTCP as TCP agent.

## B Simulation Result and Analysis

The results were collected as average values over 167 Iterations in the time interval between100 to 150 seconds. We compared the performance of M-ADTCP with ADTCP for the different metrics. In ADTCP CWL is set to constant value, where as in M-ADTCP CWL is varied based on the computed metrics.

Figure 1 show the comparison based on Average Inter Arrival Delay. The graph clearly indicates that M-ADTCP technique delivers the packets with less delay as compared to ADTCP technique.

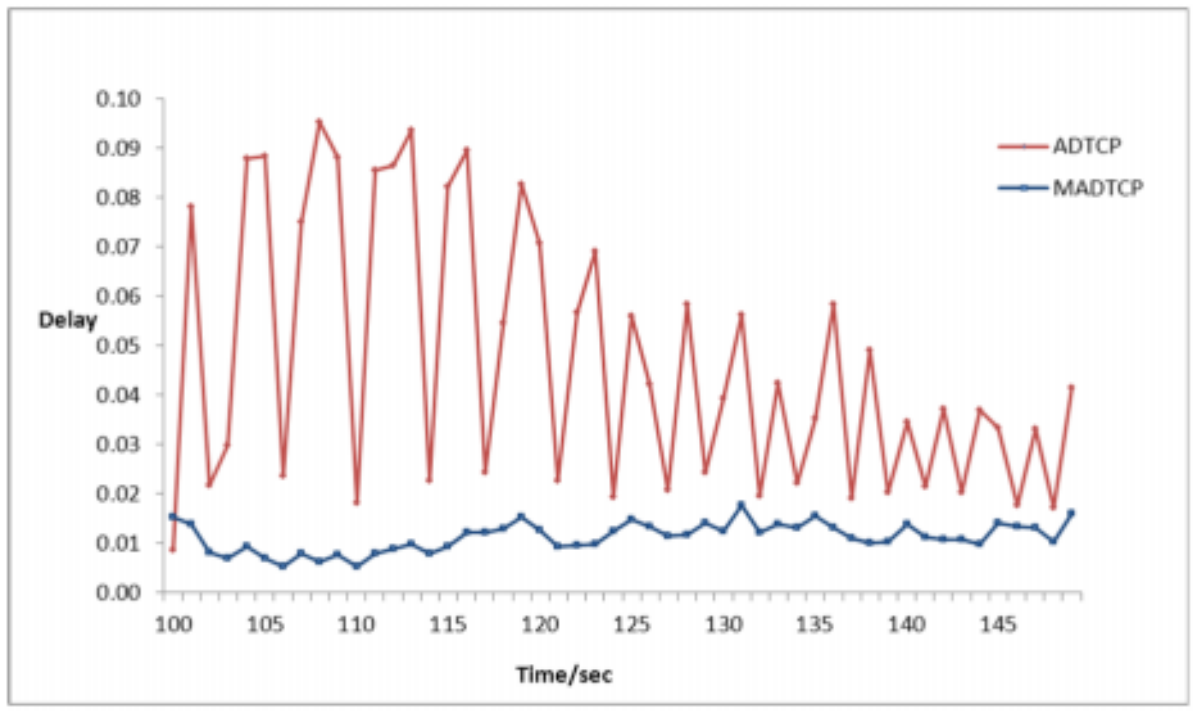

Figure 1 Comparison of Average Inter Arrival Delay between M-ADTCP and ADTCP

Figure 2 show the comparison based on Average Inter Delay Difference. The graph clearly indicates that Average Inter Delay Difference between packets is less in M-ADTCP technique as compared to ADTCP technique.

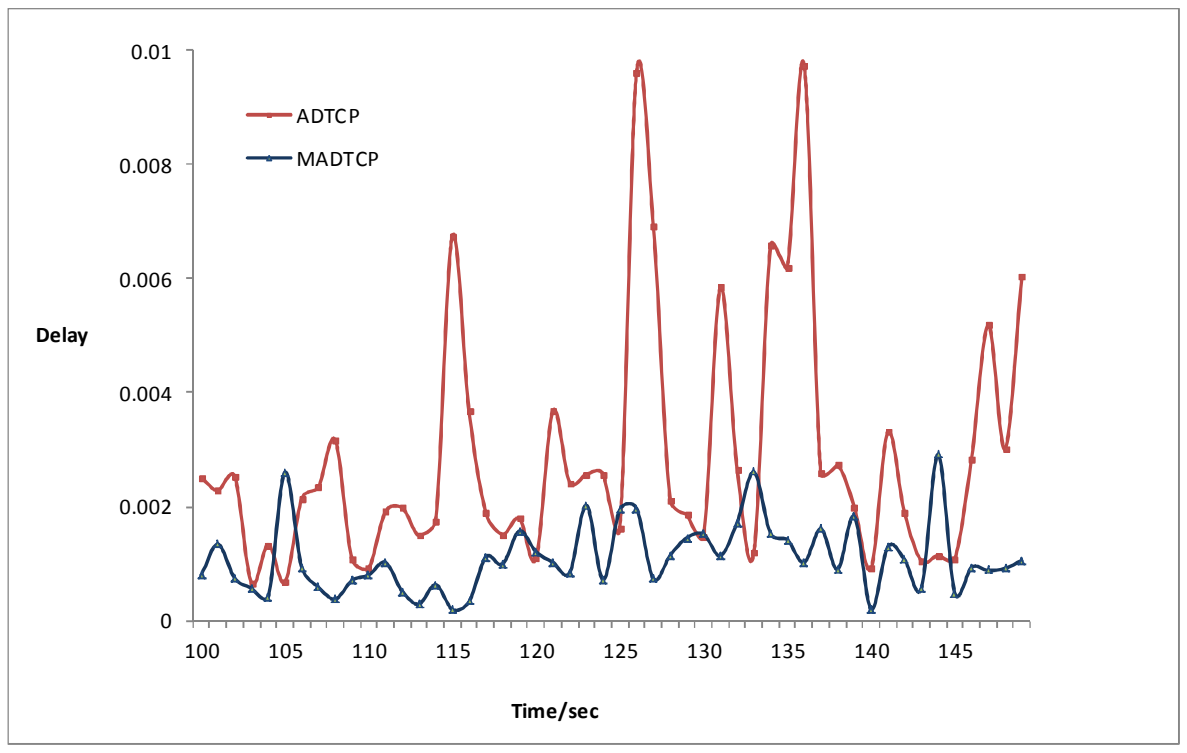

Figure 2 comparison of Average Inter Delay Difference
between M-ADTCP and ADTCP

Figure 3 show the comparison based on Packet Out of Order rate. The number of out of order packets is more in ADTCP technique as compared to M-ADTCP technique.

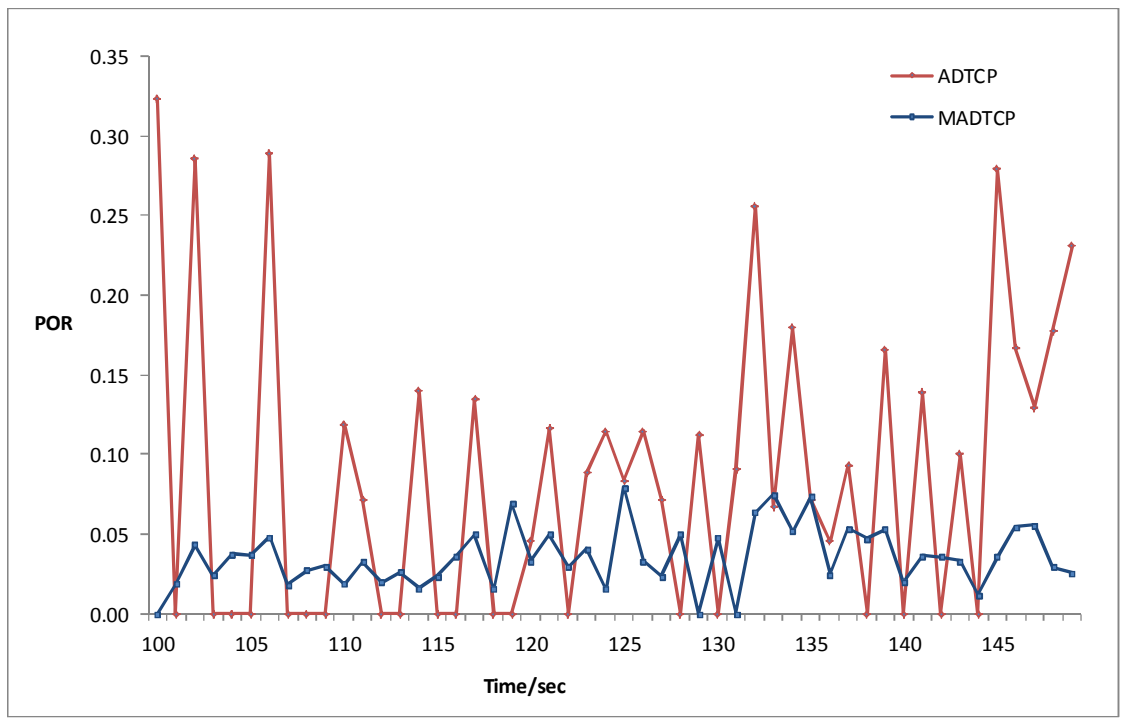

Figure 3 Comparison of Packet Out of Order rate between
M-ADTCP and ADTCP

Figure 4 shows the comparison based on Short Term Throughput metric. The graph clearly indicates that M-ADTCP technique outperforms ADTCP technique.

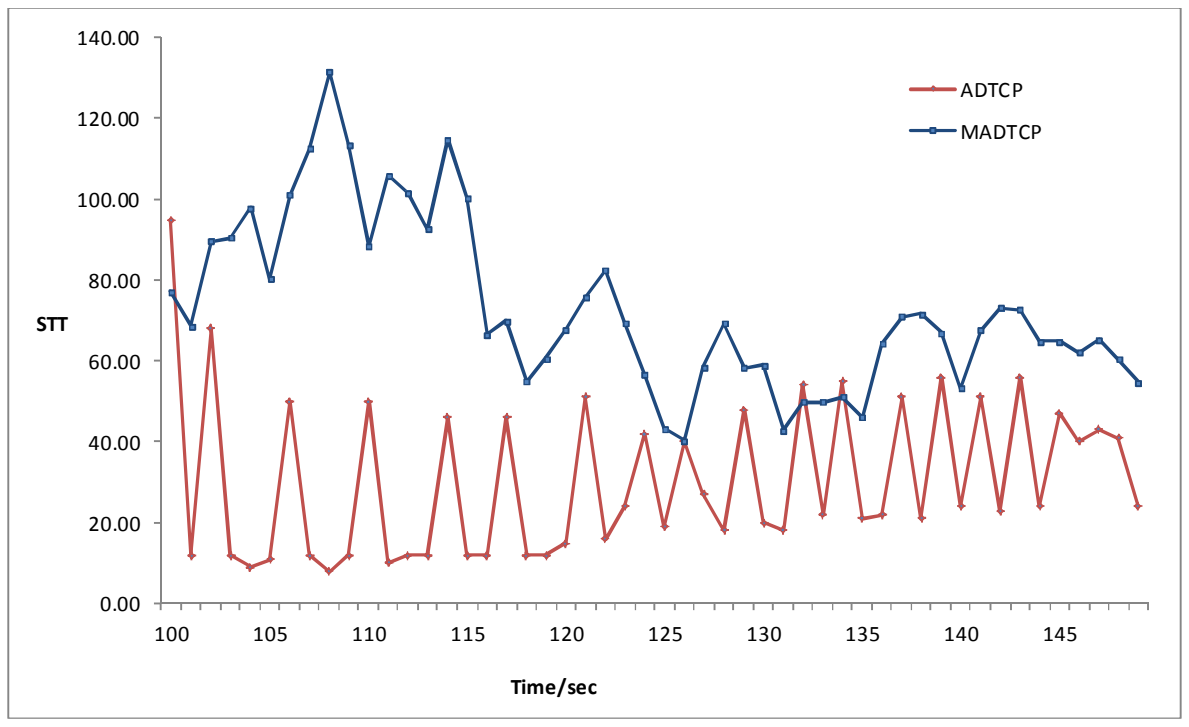

Figure 4 Comparison of Short Term Throughput between
M-ADTCP and ADTCP

# 5 CONCLUSION

Mobile Ad-hoc Networks (MANETs) have been an area of immense interest and active research over the past few years. In MANET it is very difficult to properly ascertain some of the characteristics such as channel error, loss rate, route change, congestion detection etc., as the measurement data is noisy. These limitations helped us in developing a technique which addresses those problems. From the experimental results it can be easily concluded that M-ADTCP outperforms ADTCP.

Existing TCP designed for wired network typically rely on ELFN for detecting congestion. In our approach we have adapted end-to-end measurement for congestion detection using four metrics as discussed in section 3.2. This helps M-ADTCP function well thereby increasing efficiency.

# ACKNOWLEDGEMENT

The author acknowledges Anna University Coimbatore, for the encouragement and permission to publish this paper based on the research work carried out by the author towards his Ph.d work.

The author thanks the principal of Sir MVIT, Dr.M.S.Indira for her constant encouragement and also thanks Prof. Dilip K. Sen, Head of Department Computer Science Engineering for his invaluable guidance and suggestions from time to time.

## Authors

B.C Sreenivasa is working as sr.Lecturer in the Department of computer science and Engineering Sir M visvesvaraya Institute of technology Bangalore. Published several research papers in leading journals and interested in publishing more papers in the area of wireless networks, computer network.

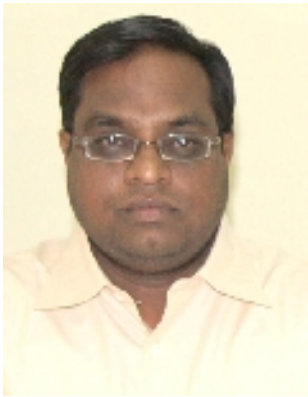

G. C. Bhanu Prakash is working as Professor in the Department of computer science and Engineering Sir M visvesvaraya Institute of technology Bangalore. Completed his PhD from MGR University Chennai in 2011. Published several research papers in leading journals and international conferences. Interested in working further in the area of wireless networks and computer networks

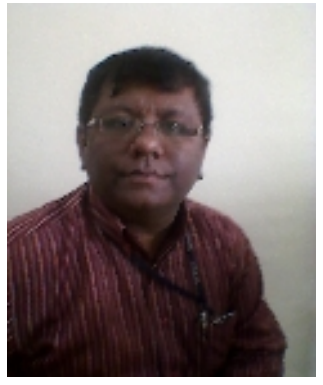